\documentclass[aps,preprint,superscriptaddress]{revtex4}
\usepackage{graphicx}
\usepackage{subfigure}
\usepackage[colorlinks,
            linkcolor=blue,
            anchorcolor=green,
            citecolor=blue
            ]{hyperref}
\usepackage{float}
\usepackage{amsmath}
\usepackage{amsfonts}
\usepackage{bm}
\usepackage{bbm}

\usepackage{amssymb}

\begin{document}
\title{Pair production in asymmetric Sauter potential well}
\author{Adiljan Sawut}
\affiliation{School of Physics Science and Technology, Xinjiang University, Urumqi, Xinjiang 830046 China}
\author{Sayipjamal Dulat \footnote{sdulat@hotmail.com}}
\affiliation{School of Physics Science and Technology, Xinjiang University, Urumqi, Xinjiang 830046 China}
\author{B. S. Xie \footnote{bsxie@bnu.edu.cn}}
\affiliation{Key Laboratory of Beam Technology of the Ministry of Education, and College of Nuclear Science and Technology, Beijing Normal University, Beijing 100875, China}
\affiliation{Beijing Radiation Center, Beijing 100875, China}
\date{\today}
\begin{abstract}
Electron-positron pair production in asymmetric Sauter potential well
is studied, where the potential well has been built as the width of the right edge fixed but the left side of the well is changeable at different values. We study the momentum spectrum, the location distribution and the total pair numbers in this  case of asymmetric potential well and compare them with the symmetric case. The relationship between created electron energy, the level energy in the bound states and the photon energy in the symmetric potential well is used to the studied problem for the created electrons in the asymmetric potential well and its validity is confirmed by this approximation. By the location distribution of the electrons we have also shown the reason why the momentum spectrum has an optimization in the asymmetric well compared with the symmetric one.
\end{abstract}
\maketitle

\section{Introduction}

The electron-positron pair production has been studied theoretically~\cite{2010The,2011Interference,2005Worldline,2006Worldline,2002Quantum,2010Schwinger} and experimentally~\cite{1985Observation,1997Search,1997Positron,1996Observation,2010Collision,2012Breit,2009Relativistic,2015The,Augustin_2014} by many authors since Sauter~\cite{Sauter1931} considered pair creation in a static electric field. In 1951~\cite{1951On}, the pair creation rate in a constant field was calculated by Schwinger and he pointed out the electric field strength for creating observable pairs, which is $E_c={10}^{16}$ \rm{V/cm}, corresponding to laser intensity of ${10}^{29}\rm{W/cm^2}$ respectively. With the rapid developments of laser technology and the chirped-pulse amplification (CPA) technology ~\cite{1985Compression,1988Amplification}, laser intensity could reach ${10}^{24}\rm{W/cm^2}$ at the present~\cite{Eli}, but still far more smaller than the critical value of Schwinger effect.
However, there is no experiment that directly achieves the conversion from energy to matter, while it is full of hope in the future.

In terms of experiment, Cowan~\cite{1985Observation} observed positive and negative electron beams in the heavy ion collision experiment in 1986.
After a decade, Ahmad~\cite{1997Search} produced a single energy positron beam through heavy ion collisions. In these experiments, however, the nuclear reaction caused by the collision of high-energy relativistic ions and the structure change of the high-Z nucleus are much greater than the effect of the Coulomb fields` superposition of two nuclei.
In general, most positrons came from a nuclear reaction rather than a direct vacuum breakdown. Also, in 1997, Burke et al.~\cite{1997Positron} used the 46.6 GeV electron beam generated by Stanford Linear Accelerator (SLAC) to collide with a ${10}^{18}W/cm^2$ laser beam, gained pair productions.
In this experiment, the non-linear Compton scattering of laser photons and electrons generates high-energy gamma photons~\cite{1996Observation} , and these high-energy gamma photons continually interact with the laser to generate positron and electron pairs, called the Breit-Wheeler process~\cite{2010Collision,2012Breit} .
In 2009, Chen~\cite{2009Relativistic} irradiated a gold target with an ultra-strong laser of ${10}^{20}W/cm^2$, and detected a positron beam of ${10}^{10}$ per sphere behind the gold target. In this experiment, laser photons and nuclear scattering generated high-energy gamma photons and these high-energy photons interact with high-Z nuclei to generate a large number of positrons and electrons, respectively, named the Bethe-Heitler process~\cite{2015The,Augustin_2014} .

In terms of theory, many theoretical ideas have been adopted for solving the non-perturbative and non-equilibrium process in pair production, such as Wentzel-Kramers-Brillouin approximation (WKB)~\cite{2010The,2011Interference} , World-line instanton technique ~\cite{2005Worldline,2006Worldline} , solving Vlasov equation ~\cite{2002Quantum} , Wigner function formalism~\cite{2010Schwinger} and so on. In this article, we discuss the pair creation problem in asymmetric potential well with computational quantum field theory (CQFT)~\cite{1999Numerical,2005Creation,2006Timing} which has had a great achievement not only with pair creation, also with Zitterbewegung~\cite{2004Relativistic} , Klein paradox~\cite{2004Klein,article} , relativistic localization problem~\cite{2004Relativistic} and so on.
Recently more interesting works were performed, which include one of ours for the effective interaction time mechanism to improve the pair numbers significantly \cite{2019Wang}. In this paper, furthermore, we study the pair creation problem in asymmetric Sauter potential well. We examine it by using the fields which include both the static asymmetric Sauter potential and the alternating field. It is found that the width of the potential well plays an important role in this electron-positron production process.

The paper is organized as follows. In Sec.\ref{the} we discuss the theoretical framework for computational quantum field theory. In Sec.\ref{multi} we study the momentum spectrum and the location distribution of the created electrons. In Sec.\ref{evolution} we compare the evolution of the number of particles created in different types of asymmetric potential well. Finally, in Sec.\ref{sum} a brief summary of the work is given.

\section{The theoretical framework for CQFT}\label{the}

As for electron and positron we use Dirac equation, we use the atomic units (a.u.) as $\hbar=m_0=e=1$ in this whole paper, as for fine structure constant $\alpha=1/c=1/137.036$, and consider a one-dimensional system along the $z$ direction for the simplicity of our calculation and the model. Here~\cite{1961An,Campbell:2017hsr}
\begin{equation}
i\partial\hat{\psi}\left({z},t\right)/\partial{t}=\lbrack{c\alpha_z{\hat{p}}_z+\beta c^2+V(z,t)}\rbrack\hat{\psi}\left(z,t\right),
\end{equation}
$\alpha_{z}$ denotes as $z$ component of the Dirac matrix, $\beta$ denotes unit Dirac matrix, and we only focus on a single spin thanks to there is no magnetic field in our one-dimensional system. In results, the four-component spinor wave function becomes two components and Dirac matrix $\alpha_{z}$ and $\beta$ are replaced by the Pauli matrix $\sigma_1$ and $\sigma_3$, where:
\begin{equation}
\sigma_1=\left(\begin{matrix}0  &  1\\1  &  0\\\end{matrix}\right),         \   \	  \	   \  	 \		\	\	\	\	\			         \sigma_3=\left(\begin{matrix}1&0\\0&-1\\\end{matrix}\right).
\end{equation}

The Hamiltonian is
\begin{equation}
{H}=c\sigma_1{\hat{p}}_z+\sigma_3 c^2+V(z,t),
\end{equation}
whereby $V\left({z},t\right)$ is the classical external scalar potential along the $z$ direction. We can express $\hat{\psi}\left(z,t\right)$ as the combine of creation and annihilation operators as follows:
\begin{equation}
\sum_p \hat{b}_p(t) W_p(z)+\sum_{n} \hat{d}_n(t) W_n(z) = \sum_p \hat{b}_p W_p(z,t)+\sum_n \hat{d}_n W_n(z,t),
\end{equation}
where $ \sum_{p(n)}\ $ stands for the summation over all states with positive and negative energy, and $ W_{p(n)}\left({z},t\right)=\langle{z}|p\left(n\right)(t)\rangle $ is the solution of the Dirac equation for the initial condition $W_{p(n)}\left({z},t=0\right)=W_{p(n)}\left({z}\right)$, where $W_{p(n)}\left({z}\right)$ is the energy eigen-function of the field-free Dirac equation. We also can express the fermion operators as:
\begin{equation}
{\hat{b}}_p(t)=\sum_{{p^\prime}}{{\hat{b}}_{p^\prime}U_{pp^\prime}\left({z},t\right)+\sum_{n^\prime}\ {\hat{d}}_{n^\prime}^\dag U_{pn^\prime}\left({z},t\right),\ \ \ \ \ \ \ \ \ \ }
\end{equation}
\begin{equation}
{\hat{d}}_n^\dag(t)=\sum_{{p^\prime}}{{\hat{b}}_{p^\prime}U_{np^\prime}\left({z},t\right)+\sum_{n^\prime}\ {\hat{d}}_{n^\prime}^\dag U_{nn^\prime}\left({z},t\right),\ \ \ \ \ \ \ \ \ \ }
\end{equation}
\begin{equation}
{\hat{b}}_p^\dag(t)=\sum_{{p^\prime}}{{\hat{b}}_{p^\prime}^\dag U_{pp^\prime}^\ast\left({z},t\right)+\sum_{n^\prime}\ {\hat{d}}_{n^\prime} U_{pn^\prime}^\ast\left({z},t\right),\ \ \ \ \ \ \ \ \ \ }
\end{equation}
\begin{equation}
{\hat{d}}_p(t)=\sum_{{p^\prime}}{{\hat{b}}_{p^\prime}^\dag U_{np^\prime}^\ast\left({z},t\right)+\sum_{n^\prime}\ {\hat{d}}_{n^\prime} U_{nn^\prime}^\ast\left({z},t\right),\ \ \ \ \ \ \ \ \ \ }
\end{equation}
where $U_{pp^\prime}\left(t\right)=\left\langle p\left|\hat{U}\left(t\right)\right|{p^\prime}\right\rangle,\ U_{pn^\prime}\left(t\right)=\left\langle p\left|\hat{U}\left(t\right)\right|{n^\prime}\right\rangle,\ U_{nn^\prime}\left(t\right)=\left\langle n\left|\hat{U}\left(t\right)\right|{n^\prime}\right\rangle,\ U_{np^\prime}\left(t\right)=\left\langle n\left|\hat{U}\left(t\right)\right|{p^\prime}\right\rangle.$

The time evolution operator of the field
$ \hat{U}\left(t\right)\equiv\hat{T}\exp{(-i\int_{0}^{t}{Hd\tau)}}$, $\hat{T} $
denotes time-order operator and due to the operators at different times may not be commutatived,
and it helps to sort the operators at different times so that the operator at the early time is classified to the right side of the operator at the late time.
The electronic portion of the field operator is defined as $ \hat{\psi}_{e}^{+}\left({z},t\right)\equiv\sum_{p}{{\hat{b}}_p\left(t\right)W_p({z})} $
so that the created electrons` spatial number density can be written as:
\begin{equation}
\rho_{e}\left({z},t\right)=\langle \text{vac}|{\hat{\psi}}_{e}^{+\dag}\left({z},t\right){\hat{\psi}}_{e}^+\left({z},t\right)|\text{vac}\rangle.
\end{equation}
The anticommutator relations $ \left\{{\hat{b}}_p,{\hat{b}}_{p^\prime}^\dag\right\}=\delta_{pp^\prime}$ and $\left\{{\hat{d}}_n,{\hat{d}}_{n^\prime}^\dag\right\}=\delta_{nn^\prime}$, the number density of the electrons can be rewritten as:
\begin{equation}
\rho_{e}\left({z},t\right)=\sum_{{n}}{\mid\sum_{{p}}{U_{pn}\left(t\right)W_p\left({z}\right)\mid^2,\ }}
\label{density}
\end{equation}
where $U_{pn}\left(t\right)$ can be computed with the split operator numerical technique~\cite{1999Numerical}. By integrating the Eq.\eqref{density}, we can obtain the total number of created electrons as:
\begin{equation}
\label{eq2}
N\left(t\right)=\int\rho_{e}\left({z},t\right)d{z}=\sum_{{p}}\sum_{{n}}{\left|U_{pn}\right|^2.\ }
\end{equation}

\section{The Multi-photon process}\label{multi}

We set the amplitude of potential well $V_1$ and alternating field $V_2$ both equal to $V_1=V_2=2c^2-10000$, and our total field is:
\begin{equation}
V\left(z,t\right)=V_1S\left(z\right)f\left(t\right)+V_2\sin{\left(\omega t\right)}S\left(z\right)\theta(t;t_0,t_0+t_1),
\end{equation}
where
\begin{equation}
 S\left(z\right)=\{[\tanh{\left(z-{D/2}\right)}/{W}_1]-[\tanh{\left(z+{D/2}\right)}/{W}_2]\}/2
 \label{shape},
 \end{equation}
and
\begin{equation}
f\left(t\right)=\sin[\pi{t}/2{t}_0]\theta(t;0,t_0)+\theta(t;t_0,t_0+t_1)+\cos[\pi({t-{t}_0-{t}_1})/2{t}_0]\theta(t;t_0+t_1,2t_0+t_1).
\label{time}
\end{equation}

Eq.\eqref{time} describes how the processes of our potential well to turn on and off, and we bring a step function $\theta\left(t;t_1,t_2\right)$ on. As demonstrated by the total field function, the oscillating field doesn't work during $(0,t_0)$ in order to reduce the trigger effect of the pair production, so we set $t_0=5/c^2$ for the well establishment of the potential well. Then we turn on the oscillating field during $(t_0,t_0+t_1)$, and $t_1=20\pi/c^2$. The total time for our calculation is $T=2t_0+t_1=(10+20\pi)c^2$ and divided into $N_t=10000$ time intervals. We have also divided the calculation space into $N_z=2048$ grid points. When we set these two parameters larger, we will get more precise results in our simulation experiment, meanwhile, the experiment time also gets longer. Therefore, they should be controlled within a reasonable range.
For the Eq.\eqref{shape}, $D$ is the width of the potential well, and $W_1,\ W_2$ are widths of the electric field at the edge. We set the $D=10\lambda_e$, where the $\lambda_e$ is the Compton wavelength. $W_1=0.3\lambda_e$ and keep stable, $W_2$ has different values, which is from $0.075\lambda_e$ to $1.5\lambda_e$, exactly quater to five times of the $W_1$, respectively. The frequency of the oscillating field is $\omega=2.1c^2$.

\begin{figure}[htbp]
\suppressfloats
\begin{center}
\includegraphics[width=\textwidth]{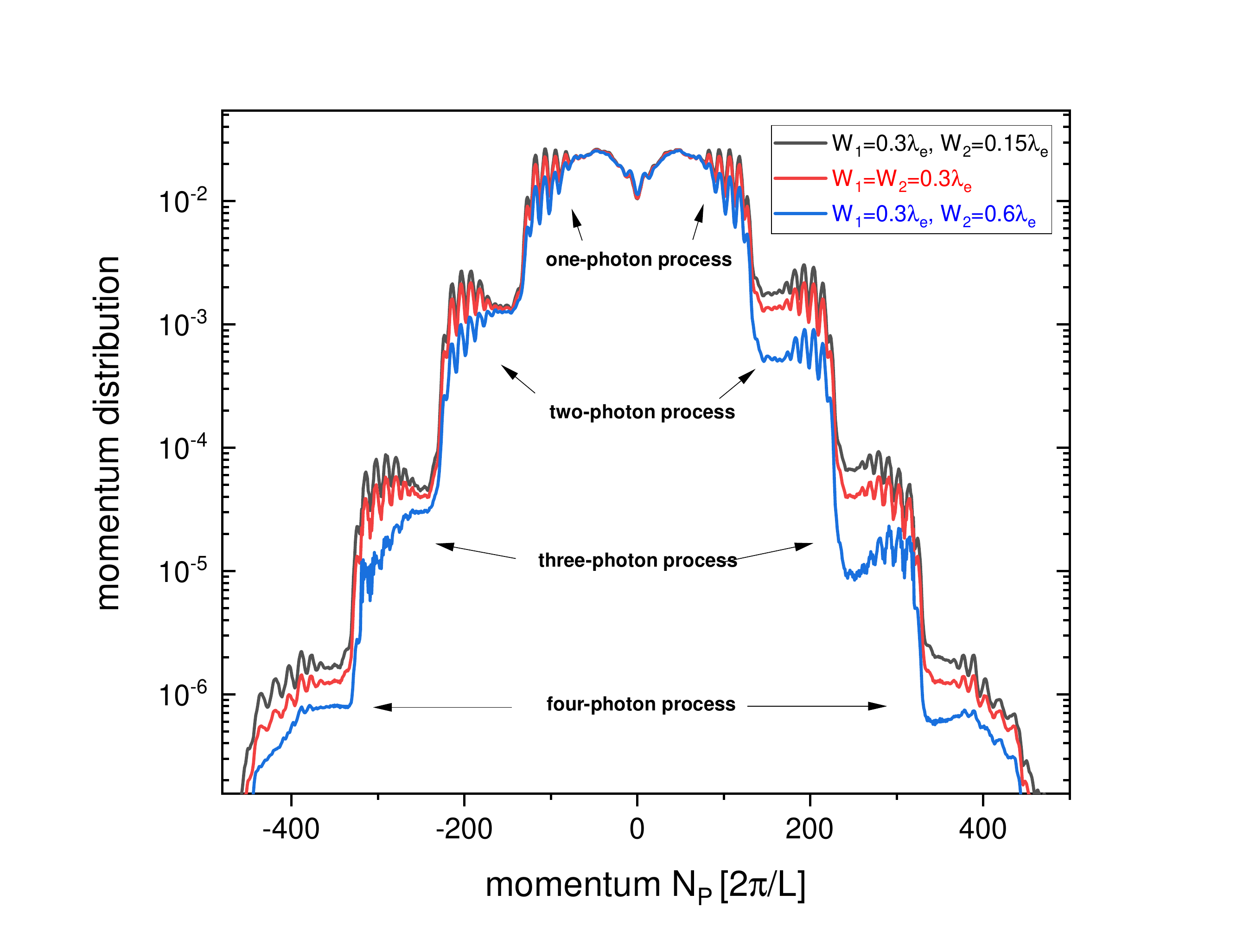}
\end{center}
\caption{Momentum spectra of the pair creation for one symmetric combined field(red), and two asymmetric combined field with different widths of the left side(black and blue spectrum). The frequency of the oscillating field is $\omega=2.1c^2$, other parameters are $N_z=2048$,\ $N_t=10000$,\ $V_1=V_2=2c^2-10000$,\  $D=10\lambda_e$, and $W$ is given in the figure.}
\label{fig:1}
\end{figure}

The FIG.\ref{fig:1} is the momentum spectrum of the pair production in three different shapes of our potential well. As we compare the black line which represents to the asymmetric potential well with $W_1=0.3\lambda_e,\ W_2=0.15\lambda_e$, and the red line which represents to the symmetric well, we can easily notice that the $N_p>0$ part has very clear changes compared with the $N_p<0$ part. The reason of this effect is that the electric field gets stronger at the left edge of our potential well as the $W_2$ gets smaller, and the electrons with positive momentum get more and more increase. On the contrary, blue line which represents to the asymmetric potential well with $W_1=0.3\lambda_e,\ W_2=0.6\lambda_e$, has clear decline in its momentum distribution compared with the symmetric potential well (red line). All this three spectra have one very common feature, which is when $N_p$ close to 0 the change is not very clear, but as $N_p>>0$ there has a marked differences between these three momentum spectra. We will discuss this in our location distribution part.

\begin{figure}[htbp]
\centering
\begin{minipage}[t]{0.48\textwidth}
\centering
\includegraphics[width=1\textwidth]{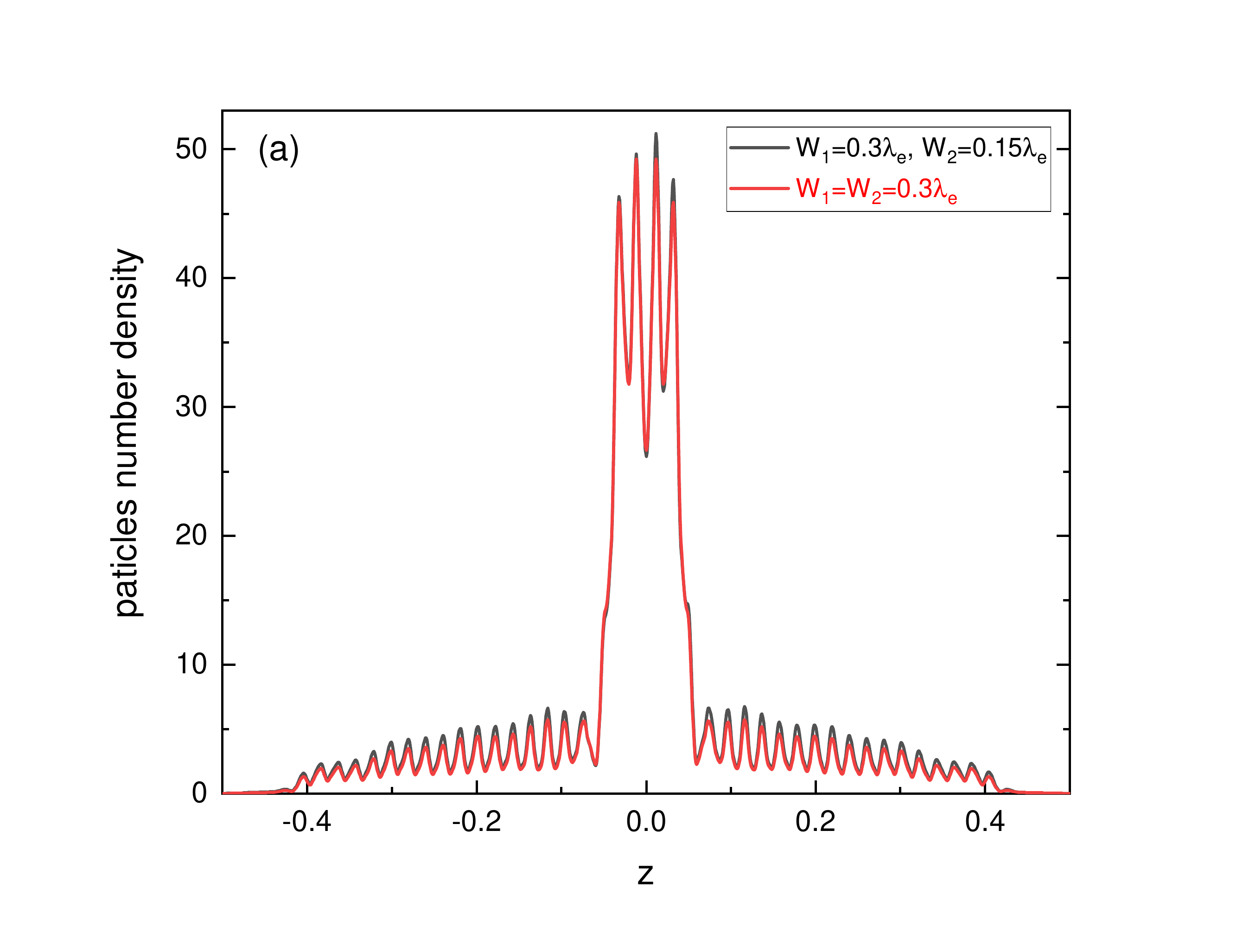}
\end{minipage}
\begin{minipage}[t]{0.48\textwidth}
\centering
\includegraphics[width=1\textwidth]{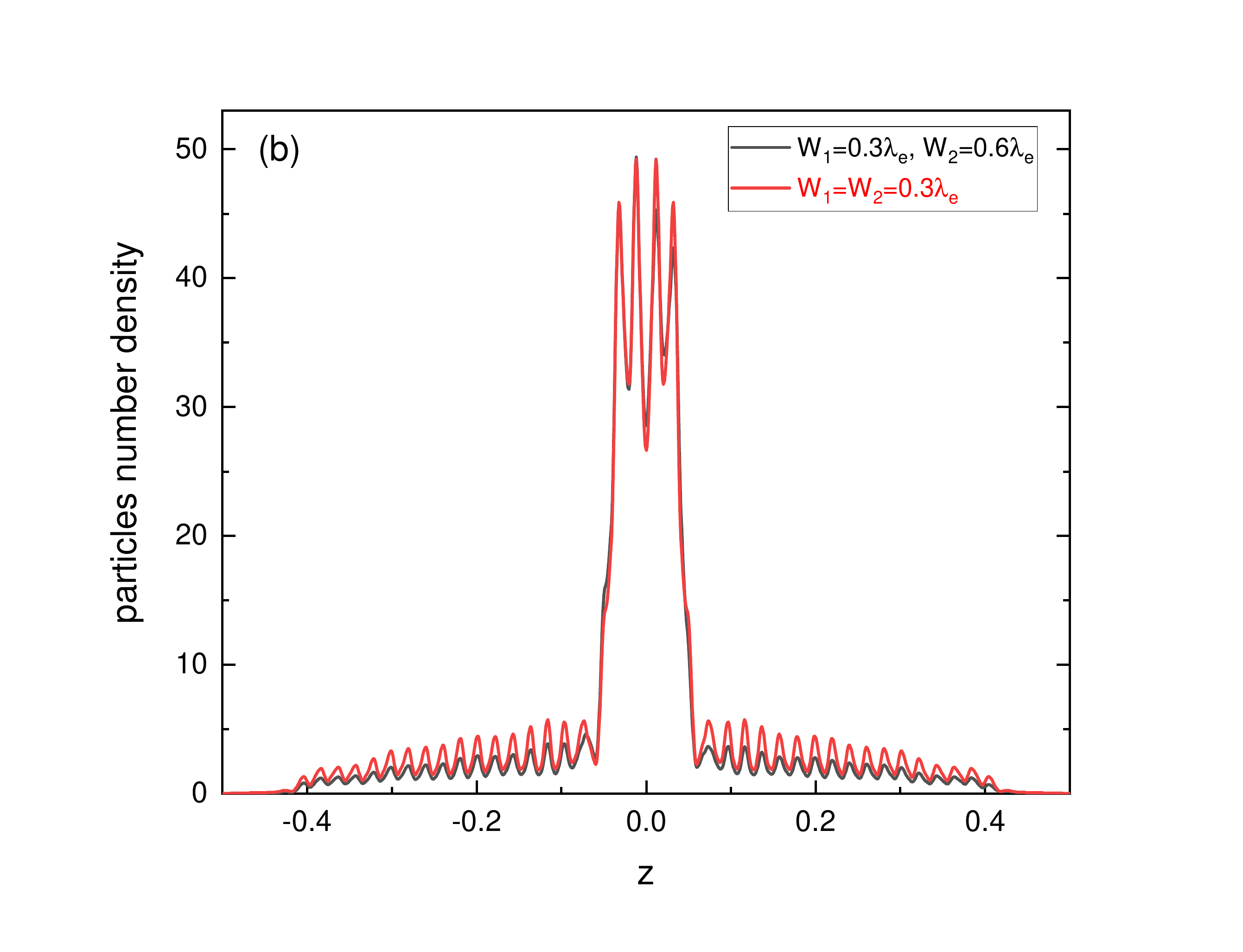}
\end{minipage}
\caption{The location distribution of the created electrons, (a) is the comparison of asymmetric potential well with $W_1=0.3\lambda_e,\ W_2=0.15\lambda_e$ (black), and symmetric well with $W_1=W_2=0.3\lambda_e$ (red), (b) is the comparison of asymmetric potential well with $W_1=0.3\lambda_e,\ W_2=0.6\lambda_e$ (black), and symmetric well with $W_1=W_2=0.3\lambda_e$ (red).}
\label{fig:2}
\end{figure}

FIG.\ref{fig:2} is the location distribution of the created electrons along the $z$ direction. In FIG.\ref{fig:2} (a), compare with the symmetric potential well, the asymmetric potential well with $W_1=0.3\lambda_e,\ W_2=0.15\lambda_e$ (black line in a) shows there are more electrons in our space. Around $z=0$ area, particle number density is little bit higher in the right side than the left side or we can say the location distribution is asymmetric when $z$ close to 0, it's very logical with our conclusions on the top. But there is also an equal amount of increase on the both side along the $\pm {z}$ direction. We can explain this phenomenon with the Fourier transformation, small-scale spatial distribution is related to large momentum and the large-scale spatial distribution is related to small momentum. As in FIG.\ref{fig:1}, the changing of width in left side has more effects to the large momentum electrons so that small-scale spatial distribution has obvious asymmetric values. The large-scale spatial distribution increase in equal amount or we can say they are symmetric, because of our asymmetric potential well has few effects to the small momentum electrons as in FIG.\ref{fig:1}. On the contrary, in FIG.\ref{fig:2} (b), there is a clear decrease on particle number density when we make the $W_2$ two times wider than the $W_1$.
	Thanks to the left side of our potential well keep changing while the right side is being stable, our spectra have optimized when width gets narrower. So in our next step, we will only discuss $W_2=0.5W_1=0.15\lambda_e$ shaped one, numerically, the detailed information about the momentum spectrum of the pair production in symmetric potential well is discussed in the reference \cite{2013Tang}. We calculate the eigenvalues of the bound states in the well with~\cite{Campbell:2017hsr}:
\begin{equation}
cp_2\cot{\left(p_2D\right)}=\frac{EV_1}{cp_1}-cp_1,
\end{equation}
whereby $p_1=\sqrt{c^2-E^2/c^2} ,\ p_2=\sqrt{(E+V_1)^2/c^2-c^2},\ \ V_1={2c}^2-10000,\ D=10\lambda_e$. This equation is symmetric in $\pm E$, so in our asymmetric well would have some uncertainties occurred.

\begin{figure}[htbp]
\centering
\begin{minipage}[t]{0.51\textwidth}
\centering
\includegraphics[width=1\textwidth]{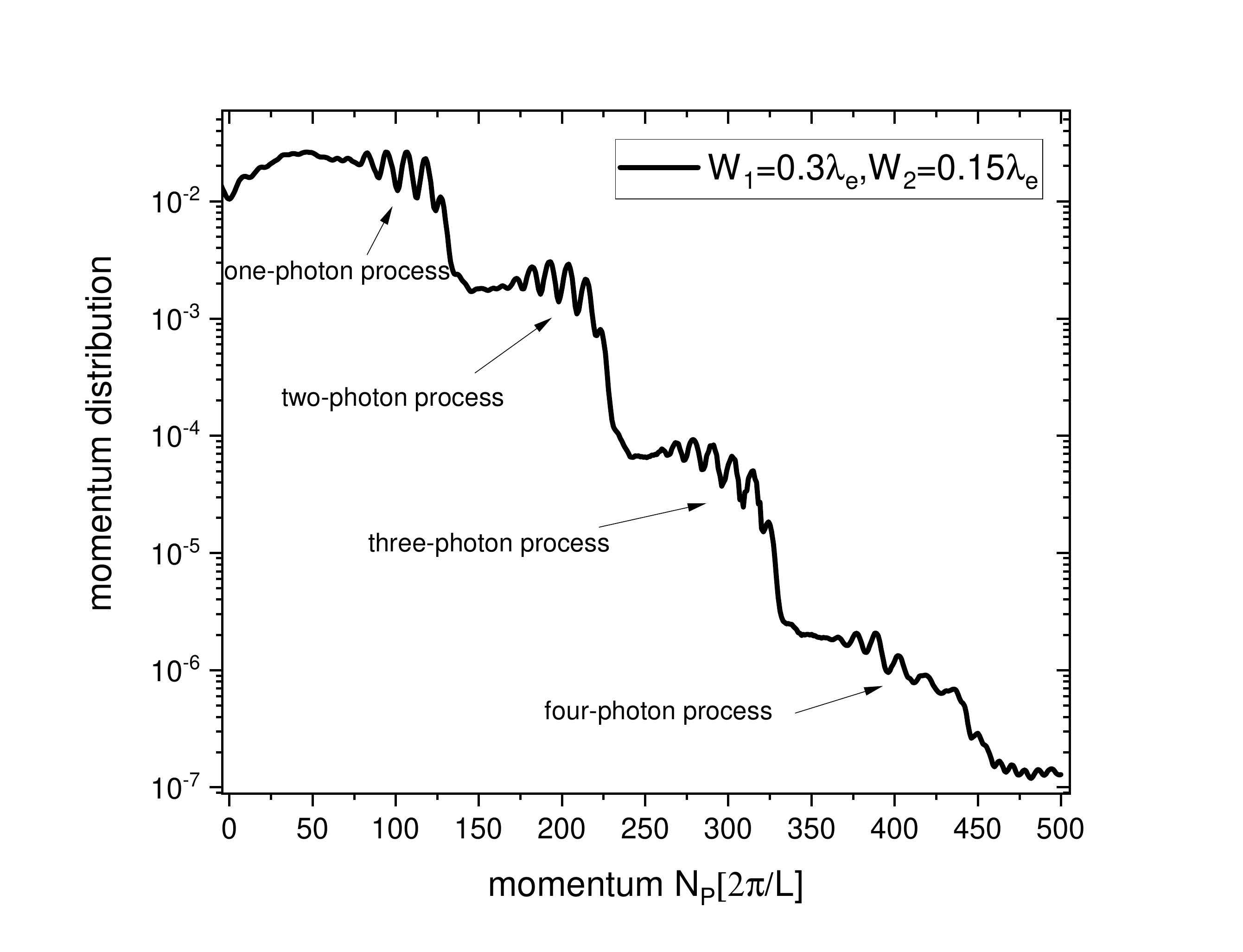}
\end{minipage}
\begin{minipage}[t]{0.48\textwidth}
\centering
\includegraphics[width=1.\textwidth]{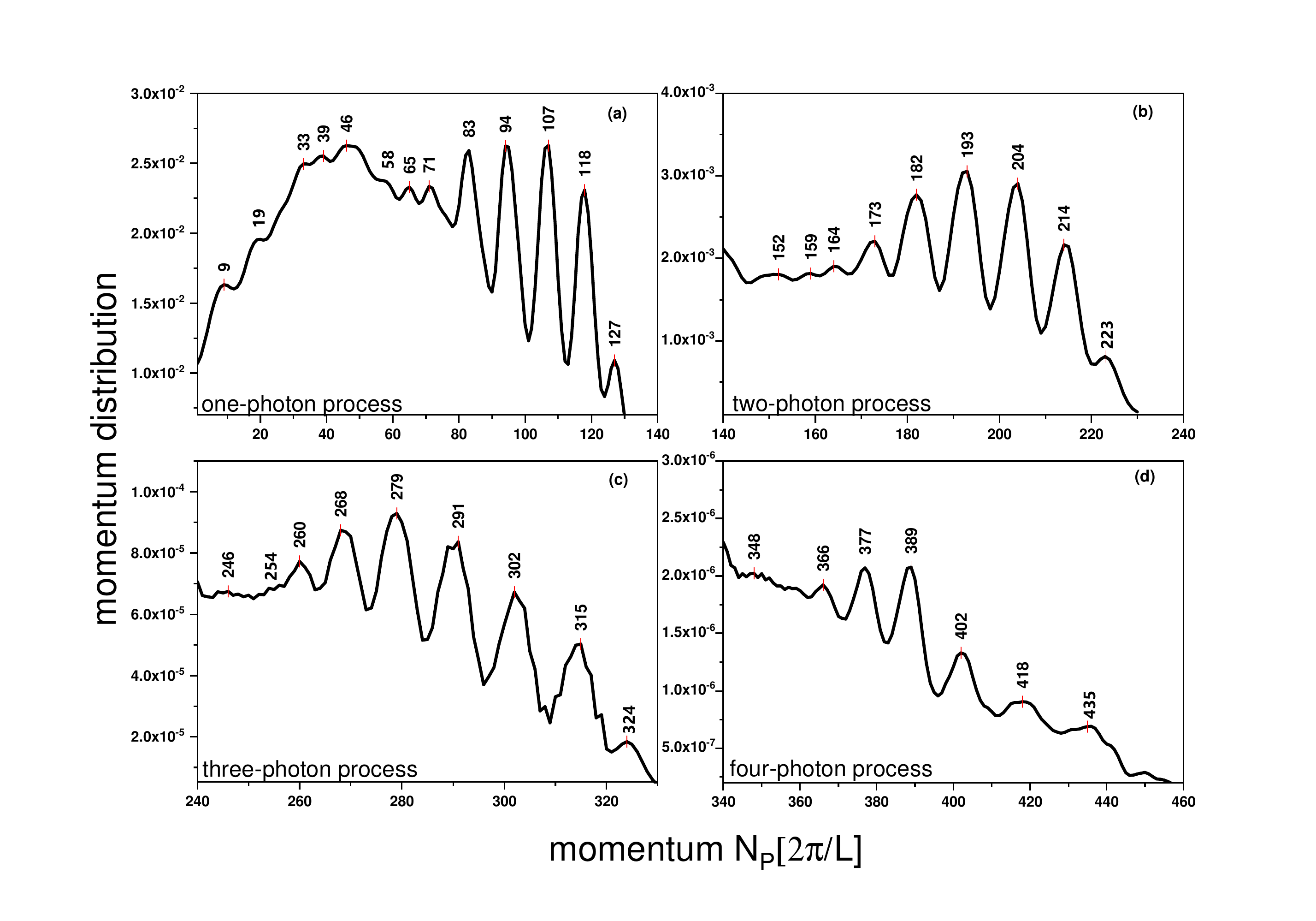}
\end{minipage}
\caption{Momentum spectrum of the asymmetric fields, the numerical grids are same as FIG.\ref{fig:1}}\label{fig:3b}
\end{figure}

We get the results as follow : $E_1=-0.4247c^2\ ,\ E_2=-0.3069c^2\ ,\ E_3=-0.1361c^2\ \ ,\ E_4=0.0680c^2\ \ ,\ E_5=0.2919c^2\ \ ,\ E_6=0.5260c^2\ \ ,\ E_7=0.7618c^2\ \ ,\ E_8=0.9978c^2$.

The relationship between peak energy in the FIG.\ref{fig:1} and the eigenvalue of the bound states in the symmetric potential well is given in the reference~\cite{2013Tang}, which is $E_{pi}=E_i+n\omega$.
We verify this relationship whether it works for our asymmetric potential well. In the FIG.\ref{fig:3b} (a), it shows the one-photon process, which means the electron jumps from the well in its bound states through absorbing one photon with frequency of $\omega=2.1c^2$. We test it with calculation of its energy,
\begin{equation}
E=\sqrt{c^2p^2+c^4},
\label{eq2}
\end{equation}
whereby $p=2N_p\pi/L$, where $L$=2.0 is the length of the numerical grid and the $N_p$ is the pick value we have marked in FIG.\ref{fig:3b}. Take an example of $N_p=58$, the eigenvalue $E_1=-0.4247c^2$ and we calculate the energy which correspond to the $N_p=58$ with Eq.\eqref{eq2}, and we get $E_{p1}=1.6637c^2$, the relationship is roughly $E_{pi}-E_i=1.6637c^2-\left(-0.4247c^2\right)=2.0884c^2\approx\omega=2.1c^2$.

We take FIG.\ref{fig:3b} (c) as an example of our calculation for multi-photon effect which is estimated that electrons in the bound states absorb three photons and become free particles. According to the values of each peak, we know $N_p=254,\ 260,\ 268,\ 279,\ 291,\ 302,\ 315,\ 324$ precisely , and each of them stand for $i=1,\ 2,\ 3,\ 4,\ 5,\ 6,\ 7,\ 8.$ And we calculate the energy of the electrons at every single peak, they can be written as $E_{N_p=254}=5.9083c^2,\ {\ E}_{N_p=260}=6.0449c^2,\ E_{N_p=268}=6.2248c^2,\ E_{N_p=279}=6.4739c^2,\ E_{N_p=291}=6.7458c^2,\ E_{N_p=302}=6.9953c^2,\ E_{N_p=315}=7.2904c^2,\ E_{N_p=324}=7.4948c^2.$

\
\begin{center}
TABLE1: The value of $E_{pi}$ and $E_i$ at the peaks
\vskip 1pt
\centering

\begin{tabular}{ |c|c|c|c|c|c|c|c|c| }
 \hline
 $i$ & 1 & 2 & 3 & 4 & 5 & 6 & 7 & 8\\
\hline
 $E_i$ & $-0.4247c^2$ & $-0.3069c^2$ & $-0.1361c^2$ & $0.0680c^2$ & $0.2919c^2$ & $0.5260c^2$ & $0.7618c^2$ & $0.9978c^2$\\
\hline
 $N_p$ & 254 & 260 & 268 & 279 & 291 & 302 & 315 & 324\\
\hline
 $E_{pi}$ & $5.9083c^2$ & $6.0449c^2$ & $6.2248c^2$ & $6.4739c^2$ & $6.7458c^2$ & $6.9953c^2$ & $7.2904c^2$ & $7.4948c^2$\\
 \hline
\end{tabular}
\end{center}
\

With very simple calculation we verify that $E_{pi}=E_i+3\omega$, also indicate our estimation is fully correct.

The changing of the momentum spectrum as we turnover the widths at the two side of the field is predictable. We have just compared two groups of data, which are $W_1=0.3\lambda_e,W_2=0.6\lambda_e$ and $W_1=0.6\lambda_e,W_2=0.3\lambda_e$ respectively. Obviously, the momentum distribution of positive and negative $N_p$ has transferred its corresponding values as we turnover the widths at each edge.

\section{Evolution of the electron number}\label{evolution}
In this section we study the pair production in subcritical asymmetric potential well with a subcritical oscillating field, but with different widths of electric fields as we have considered in the last section. We set $V_1=V_2={2c}^2-10000$, and $\omega=2.1c^2$.
\begin{figure}[htbp]
\suppressfloats
\begin{center}
\includegraphics[width=1\textwidth]{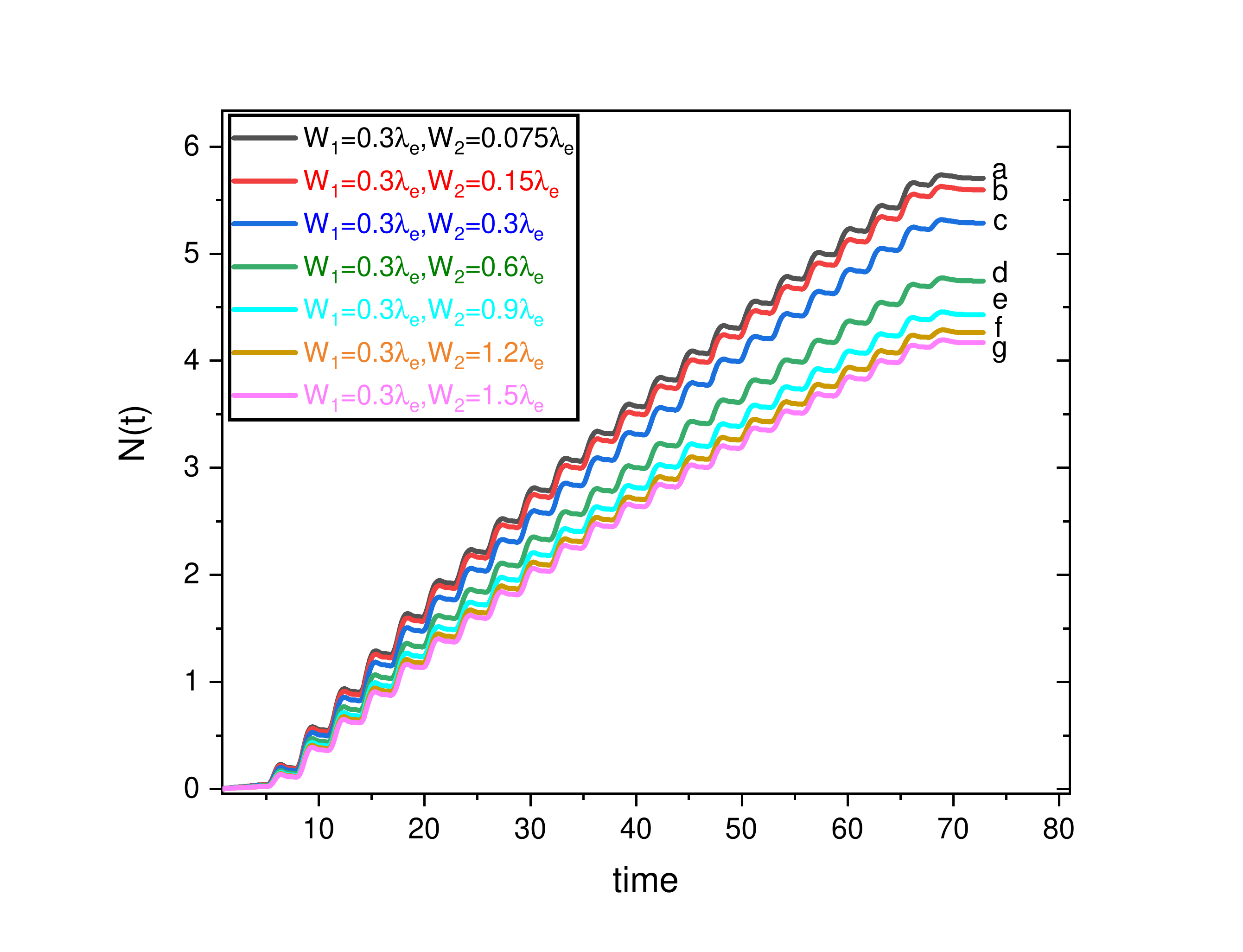}
\end{center}
\caption{The total number $N(t)$ for the created electrons in the different well with different widths given very clearly in the figure. The parameters are $\omega=2.1c^2,\ N_z=2048,\ N_t=10000,\ V_1=V_2=2c^2-10000,\ D=10\lambda_e,\ L=2.0.$ }\label{fig:5}
\end{figure}

In FIG.\ref{fig:5}. All the curves have only one difference, other parameters are the same. The right width of the electric field at the edge of the potential well is stable($W_1=0.3\lambda_e$), and we keep changing the left side from 0.25 time to 5 times of the right side which are namely $W_2=0.075\lambda_e,0.15\lambda_e,0.3\lambda_e,0.6\lambda_e,0.9\lambda_e,1.2\lambda_e,1.5\lambda_e$. The curve $c$ represents to the symmetric potential well, other six curves represent to the asymmetric potential well.
	We prefer to take curve $c$ as a comparison with other six curves. As shown in FIG.\ref{fig:5}, when we keep the width of right side of the potential well stable, make the left side wider, the rate of the electron creation would be slower and slower over time, as in curves $d$, $e$, $f$ and $g$, and close to one very specific value as in curve $g$, which approximately equals to the value of a one-sided potential well. The shape of this potential-well can be written as follow:
\begin{equation}
V\left(z,t\right)=V_1S\left(z\right)f\left(t\right)+V_2\sin{\left(\omega t\right)}S\left(z\right)\theta(t;t_0,t_0+t_1),
\end{equation}
where   $S\left(z\right)=({1+\tanh[{z/W}]})/2$, it is shown that when one side of the potential well is enough wider, and it doesn't have any effect to the whole system, which we call a critical value of the electric field that can hardly effect to the increasing of the pair production.

We realize that compared with the curve $c$, the yields of the pair production has increased in curve $a$ and curve $b$, as the curve $a$ represents to the yields of the electrons in the potential well with $W_1=0.3\lambda_e,$ $W_2=0.075\lambda_e$ and the curve $b$ represents to the yields of the electrons in the potential well with $W_1=0.3\lambda_e,$ $W_2=0.15\lambda_e$. Surely it is logical due to the intensity of the electric field would be stronger as the width gets narrower. As shown in the FIG.\ref{fig:5}, increase of the yields is evident from $W_2=0.3\lambda_e$ to $W_2=0.15\lambda_e$, and it has very weak effect as we make the $W_2$ narrower, $W_2=0.075\lambda_e$, precisely. That is approximately the maximum value of the yields in our asymmetric potential well.

\section{Summary}\label{sum}

We have compared the momentum spectrum, location distribution and the yields of the pair production in different shape of potential well, in other word, symmetric and asymmetric potential well. As for the momentum spectrum in symmetric case, it had obtained a simple law which is $E_{pi}=E_i+n\omega,$  we find this relationship is also useful for the asymmetric potential well after calculation. It means electrons in the bound states created in the well, escape from their positions after absorbing one or more photons. The momentum distribution of the electrons with positive momentum gets obvious increase than the negative momentum electrons when the left side of potential well becomes narrower. On the contrary, the momentum distribution of the electron with positive momentum gets obvious decrease than the negative momentum electrons when the left side of potential well becomes wider, as expected. The reason for this effects is explicated by the location distribution as shown in FIG.\ref{fig:2}.
	
For the yields of the pair production, the growth of $N(t)$ gets slower increase when we make one of the width of the potential well wider, because of the electric intensity gets weaker correspondingly, and close to the value of one-sided potential well. The growth of $N(t)$ increases much significantly as the width gets narrower. In general, because of the multi-photon effects, the pair production always keeps increasing in any type of our asymmetric potential well.

\begin{acknowledgments}
\noindent

We would like to thank Binbing Wu and Suo Tang for useful discussions. The work was supported by the National Natural Science Foundation of China (NSFC) under Grants No. 11875007 , No. 11935008 and No. 11965020. The numerical simulation was carried out at CTP office in Xinjiang University.

\end{acknowledgments}


\begin{thebibliography}{99}\suppressfloats

\bibitem{2010The}
C. K. Dumlu and G. V. Dunne, Phys. Rev. Lett. \textbf{104}, 250402 (2010).

\bibitem{2011Interference}
C. K. Dumlu and G. V. Dunne, Phys. Rev. D \textbf{83}, 220 (2011).

\bibitem{2005Worldline}
G. Dunne and C. Schubert, Phys. Rev. D \textbf{72}, 141 (2005).

\bibitem{2006Worldline}
G. Dunne, Q. H. Wang, H. Gies, and C. Schubert, Phys. Rev. D \textbf{73},
253 (2006).

\bibitem{2002Quantum}
C. D. Roberts, S. M. Schmidt, and D. V. Vinnik, Phys. Rev. Lett. \textbf{89}, 153901 (2002).

\bibitem{2010Schwinger}
F. Hebenstreit, R. Alkofer, and H. Gies, Phys. Rev. D \textbf{82}, 265 (2010).

\bibitem{1985Observation}
T. Cowan \emph{et al.}, Phys. Rev. Lett. \textbf{56}, 444 (1985).

\bibitem{1997Search}
I. Ahmad \emph{et al.}, Phys. Rev. Lett. \textbf{78}, 618 (1997).

\bibitem{1997Positron}
D. L. Burke \emph{et al.}, Phys. Rev. Lett. \textbf{79}, 1626 (1997).

\bibitem{1996Observation}
C. Bula, K. T. Mcdonald, and E. J. Prebys, Phys. Rev. Lett. \textbf{76}, 3116 (1996).

\bibitem{2010Collision}
G. Breit and J. A. Wheeler, Phys Rev \textbf{46}, 1087 (1934).

\bibitem{2012Breit}
K. Krajewska and J. Z. Kaminski, Phys. Rev. A \textbf{86}, 7877 (2012).

\bibitem{2009Relativistic}
H. Chen \emph{et al.}, Phys. Rev. Lett. \textbf{102}, 105001 (2009).

\bibitem{2015The}
W. Heitler, Inter. Jour. Quan. Chem. \textbf{64}, 735 (2015).

\bibitem{Augustin_2014}
S. Augustin and C. Mller, Jour. Phys.: Conf. Series \textbf{497}, 012020 (2014).

\bibitem{Sauter1931}
F. Sauter, Z. Phys. \textbf{69}, 742 (1931).

\bibitem{1951On}
J. Schwinger, Phys. Rev. \textbf{82}, 664 (1951).

\bibitem{1985Compression}
D. Strickland and G. Mourou, Opt. Commun. \textbf{55}, 447 (1985).

\bibitem{1988Amplification}
P. Maine and G. Mourou, Opt. Lett. \textbf{13}, 467 (1988).

\bibitem{Eli}
https://www.eli-beams.eu/

\bibitem{1999Numerical}
J. W. Braun, Q. Su, and R. Grobe, Phys. Rev. A \textbf{59}, 604 (1999).

\bibitem{2005Creation}
P. Krekora, K. Cooley, Q. Su, and R. Grobe, Phys. Rev. Lett. \textbf{95}, 070403 (2005).

\bibitem{2006Timing}
C. C. Gerry, Q. Su, and R. Grobe, Phys. Rev. A \textbf{74}, 044103 (2006).

\bibitem{2004Relativistic}
P. Krekora, Q. Su, and R. Grobe,  Phys. Rev. Lett. \textbf{93}, 043004 (2004).

\bibitem{2004Klein}
P. Krekora, Q. Su, and R. Grobe,  Phys. Rev. Lett. \textbf{92}, 040406 (2004).

\bibitem{article}
R. Alkofer,  \emph{et al.}, Phys. Rev. Lett. \textbf{87},
193902 (2001).

\bibitem{2019Wang}
L. Wang, B. Wu, and B. S. Xie, Phys. Rev. A \textbf{100}, 022127 (2019).

\bibitem{1961An}
S. S. Schweber and J. C. Polkinghorne, Physics Today \textbf{15}, 66 (1961).

\bibitem{Campbell:2017hsr}
W.Greiner, \emph{Relativistic Quantum Mechanics: Wave Equations}, 3rd ed. (Springer-Verlag, Berlin, 2000)
\bibitem{2013Tang}
S. Tang, B. S. Xie, D. Lu, H. Y. Wang, L. B. Fu, and J. Liu, Phys. Rev. A \textbf{88}, 012106 (2013).


\end{thebibliography}

\end{document}